\documentclass[12pt]{iopart}

\usepackage{iopams}  
\usepackage{graphicx,xcolor}

\newcommand{\opa}{\hat{a}}

\newcommand{\pare}[1]{\left(#1\right)}
\newcommand{\beqn}{\begin{eqnarray}}
\newcommand{\eeqn}{\end{eqnarray}}
\newcommand{\ket}[1]{\left\vert#1\right\rangle}

\newcommand{\proj}[2]{\left\vert#1\rangle\langle#2\right\vert}

\newcommand{\be}{\begin{equation}}
\newcommand{\ee}{\end{equation}}
\newcommand{\ba}{\begin{array}}
\newcommand{\ea}{\end{array}}
\newcommand{\bea}{\begin{eqnarray}}
\newcommand{\eea}{\end{eqnarray}}

\newcommand{\bqa}{\begin{eqnarray}}
\newcommand{\eqa}{\end{eqnarray}}
\begin{document}

\title[Quantum Simulations of Relativistic Quantum Physics in Circuit QED]{Quantum Simulations of Relativistic Quantum Physics in Circuit QED}

\author{J S Pedernales$^{1}$, R Di Candia$^{1}$, D Ballester$^{2,1}$ and E Solano$^{1,3}$}

\address{$^1$Department of Physical Chemistry, University of the Basque Country UPV/EHU, Apartado 644, 48080 Bilbao, Spain\\
$^2$Department of Physics and Astronomy, University College London, Gower Street, London WC1E 6BT, United Kingdom\\
$^3$IKERBASQUE, Basque Foundation for Science, Alameda Urquijo 36, 48011 Bilbao, Spain}

\begin{abstract}
We present a scheme for simulating relativistic quantum physics in circuit quantum electrodynamics. By using three classical microwave drives, we show that a superconducting qubit strongly-coupled to a resonator field mode can be used to simulate the dynamics of the Dirac equation and Klein paradox in all regimes. Using the same setup we also propose the implementation of the Foldy-Wouthuysen canonical transformation, after which the time derivative of the position operator becomes a constant of the motion. 
\end{abstract}


\maketitle

\section{Introduction}

R. P. Feynman is credited for introducing the idea of a quantum simulator \cite{Feynman}. At his keynote speech during the 1st Conference on Physics and Computers in 1981, he claimed that because the laws of Nature are not those of classical mechanics, it would be better to build a quantum simulation of a physical problem under the same quantum mechanical laws. For the following ten years, his contribution was somewhat hindered by the work of Deutsch \cite{Deutsch}, who showed the existence of a universal quantum computer, a quantum mechanical generalization of the classical Turing machine using quantum logic gates and algorithms.

The concept of quantum simulators will reappear later in 1996, when Lloyd presented a general proof for Feynman's conjecture \cite{Lloyd}. Quantum simulators can surpass the capabilities of classical ones, where the complexity of a problem scales exponentially. Lloyd argues \cite{Lloyd} that whereas the simulation of a general 40 spin$-1/2$ particle system would be enough to outperform existing classical computers, the factorization of a 100-digit number with Shor's algorithm would need of thousands of qubits to become nontrivial. Therefore, it is likely that the first quantum device to beat classical computers would be a quantum simulator. But there is another reason why quantum simulators are attracting so much attention lately. They provide powerful analogies to understand the physics of most varied type of systems one can think of, including condensed-matter physics, high-energy physics, cosmology and quantum chemistry \cite{Nori, Lanyon}. When compared to the original systems, quantum simulators show an unprecedented degree of controllability over all physical parameter regimes.

In this work, we show how the physics of relativistic quantum mechanics can be simulated in a very different setting. That of circuit quantum electrodynamics (cQED) and superconducting qubits. We will focus our attention on the fundamental equation in relativistic quantum physics, the Dirac equation. Proposed by Dirac himself in an attempt to combine quantum mechanics and special relativity in 1928 \cite{PAMD}, this equation is the first one to describe elementary spin$-1/2$ particles---such as the electron---and the existence of antimatter. Here, we restrict our discussion to the simplest case of 1+1 dimensions for which the algebra of the Dirac spinors can be reduced to that of standard Pauli matrices \cite{Thaller}. In the standard representation \cite{Thaller}, the 1+1 Dirac equation can be written as
\begin{eqnarray}
i\hbar \frac{d \psi}{dt}
=  \pare{ mc^2 \sigma_z    +   c \hat{P}  \sigma_y   } \psi , \label{Dirac}
\end{eqnarray}
where $\sigma_y$ and $\sigma_z$ are Pauli matrices, $\hat{P}$ represents the momentum of a Dirac particle of mass $m$ and $c$ the speed of light. Soon after its introduction, Schr\"odinger realised that there is something odd in the solution of this equation \cite{Schroe}. The position for a free Dirac particle can be integrated to give \cite{Lock}
\begin{eqnarray}
\hat{x}(t) = \hat{x}_0 + c^2 \hat{P} {\cal H}^{-1} t + \hat{Z}(t) ,    \label{evolx}
\end{eqnarray}
with $ {\cal H}= mc^2 \sigma_z    +   c \hat{P}  \sigma_y  $ and
\begin{eqnarray}
\hat{Z}(t) = \frac{i}{2} \hbar c \pare{  \sigma_y -  c \hat{P} {\cal H}^{-1} }   {\cal H}^{-1}    \pare{\exp\left(- i 2 {\cal H} t/\hbar \right)  - 1} .    \label{evolZ}
\end{eqnarray}
This third term appears because the time derivative of the position operator is not a constant of motion, i.e. $ [  \dot{ \hat{x}  } , {\cal H} ] \neq 0$, even for a free particle. Given a general initial wavepacket, the expectation value $\langle  \hat{Z}(t) \rangle$ shows an oscillating pattern with amplitude $\sim \hbar/mc$ and frequency $\sim 2mc^2/\hbar$ that is known as {\it Zitterbewegung} or {\it jittering motion}~\cite{Lock}. This behaviour stems from the interference between positive and negative energy components. Indeed, it is well established that initial states with only positive or negative components will not show any {\it Zitterbewegung} \cite{Thaller, Lock}. In spite of the appealing physics behind the Dirac equation, there has never been an experiment to test the existence of such an effect. In fact, for relativistic electrons, one can estimate a {\it Zitterbewegung} with an amplitude of about $10^{-4}$ nm and a frequency of $10^{21}$ Hz, making its eventual detection very demanding.

There is another interesting property of the Dirac equation related to the behaviour of relativistic Dirac particles under the effect of an external scalar potential,
\begin{eqnarray}
i\hbar \frac{d \psi}{dt}
=  \left\{  mc^2 \sigma_z    +   c \hat{P}  \sigma_y   + \Phi (\hat{X})   \right\} \psi . \label{Klein}
\end{eqnarray}
In case of a nonrelativistic particle impinging against a semi-infinite step potential $\Phi (\hat{X})$, we know that if the kinetic energy of the particle is smaller than the potential height, no propagating solutions exist at the other side of the potential barrier. In other words, the particle penetrates slightly, but it will be eventually reflected. Klein found out that Eq.~(\ref{Klein}) for a relativistic Dirac particle allows for propagating solutions beyond the potential barrier \cite{Klein}. And in the case of an infinitely high potential, the tunneling occurs with unit probability. More precisely, while for a small potential $\Phi < E - mc^2$ the particle with kinetic energy $E$ will be mostly transmitted, for a larger one $ E - mc^2 < \Phi  < E + mc^2 $ it will be reflected. However, making even larger the potential $ \Phi > E + mc^2 $ the particle will propagate through the barrier. To explain this apparent paradox, one can make use of the idea of particle-antiparticle creation by the potential \cite{Thaller}. A particle can go through the potential by turning into its antiparticle.

Although the Dirac equation is considered a milestone of relativistic quantum physics, playing a crucial role towards the later development of quantum field theories, in the last decades it has driven a renewed interest in different playgrounds. Examples include: superconductivity, where Bogoliubov quasi-particles can exhibit {\it Zitterbewegung}~\cite{Lurie}; semiconductor physics, with a close analogy between the $\mathbf{k}\cdot\mathbf{p}$ theory of energy bands in narrow-gap semiconductors and the Dirac equation \cite{Zawadzki}; the physics of graphene, characterised by quasi-electrons behaving like relativistic Dirac particles in two dimensions \cite{graphene}; a proposal for generating relativistic scattering in optical latices \cite{Weitz11}; photonic arrays with implementation of the quantum Rabi model and the Dirac equation \cite{Osellame12, Szameit10}; and the physics of ion traps, where the vibrational and internal degrees of freedom of the ions can be made to follow relativistic quantum dynamics \cite{Dirac,DiracExp,KleinP,KleinPExp,Lucas}, or even behave as interacting bosons and fermions \cite{Casanova12, Mezzacapo12}. In the spirit of quantum simulations, the latter represents an important advancement in terms of full quantum controllability of the dynamics and the retrieval of information through measurement.

Here, we present a scheme for performing quantum simulations of relativistic quantum mechanics using state-of-the-art cQED technology. This is motivated by two reasons. First, since 2004 \cite{Blais,Wallraff2004}, cQED has become the quantum platform with the most promising perspectives in terms of scalability and coherence. Second, there are crucial physical differences with respect to any previous implementation of the Dirac equation and Klein paradox. In our case, the physics of a relativistic spin$-1/2$ particle is simulated by the interaction of two degrees of freedom of different physical systems, i.e., a standing wave in a superconducting resonator and the two lowest levels of a superconducting qubit, where none of them move at all. The position and momentum of the Dirac particle are then encoded in its phase-space or field quadrature representation. This opens up new possibilities for combining intracavity fields with propagating quantum microwaves \cite{Menzel,Wallraff,Itinerant} in scalabale quantum network architectures \cite{Leib12} with delocalised and/or sequential interactions.

\section{General derivation}

For our protocol, we require a two-level superconducting qubit, as a flux qubit~\cite{flux}, working at its symmetry point, strongly coupled to a single electromagnetic field mode of the resonator. This interaction will be described by the Jaynes-Cummings model (JCM) \cite{JC,Blais,Wallraff2004}. In addition, we introduce three external classical microwave drivings, two transversal to the resonator \cite{qsusc}, coupling only to the qubit, and one longitudinal that only sees the resonator mode. The time dependent Hamiltonian of the complete system is given by
\begin{eqnarray}
  {\cal H} &=&  \frac{\hbar \omega_q}{2} \sigma_z +\hbar  \omega \opa^\dag \opa -\hbar  g \pare{\sigma^\dag \opa + \sigma \opa^\dag}   - \hbar \Omega  \pare{ e^{ i \pare{ \omega t +\varphi} } \sigma + e^{-i \pare{ \omega t +\varphi } } \sigma^\dag}  \nonumber \\ 
  & &  - \hbar  \lambda   \pare{ e^{i  \pare{ \nu t +\varphi }} \sigma + e^{-i  \pare{ \nu t +\varphi } } \sigma^\dag} + \hbar \xi \pare{  e^{i \omega t} \opa + e^{-i \omega t} \opa^\dag  }  . \label{HamilDriv}
\end{eqnarray}
with $\sigma_y=i \sigma - i \sigma^\dag=i \proj{g}{e} - i \proj{e}{g}$ and $\sigma_z= \proj{e}{e} -  \proj{g}{g}$, $\ket{g}$ and $\ket{e}$ being the ground and excited states of the qubit. Here, $\hbar\omega$ and $\hbar\omega_q$ stand for the photon energy and qubit energy splitting, whereas $g$ is the coupling constant. Likewise, the two orthogonal drivings have real amplitude $\Omega$, $\lambda$, phase $\varphi$, and frequencies $\omega$, $\nu$. Finally, the longitudinal driving is characterised by an amplitude $\xi$ and a frequency $\omega$. Note that while we have chosen two of the drivings to be resonant with the resonator field mode, the other parameters will be set later on. For simplicity in our presentation, we will also assume that $\omega_q=\omega$, i.e. qubit and resonator field interact resonantly as well.

Our derivation consists of two straightforward transformations. First, Hamiltonian in Eq.~(\ref{HamilDriv}) can be simplified using the reference frame rotating with the resonator frequency $\omega$
\begin{eqnarray}
{\cal H}^{L_1} &=&  - \hbar g  \pare{\sigma^\dag \opa +\sigma \opa^\dag}  - \hbar \Omega \pare{ e^{ i \varphi } \sigma + e^{-i \varphi } \sigma^\dag} + \hbar \xi \pare{ \opa + \opa^\dag}    \nonumber \\ & & - \hbar  \lambda   \pare{ e^{i \left[ (\nu-\omega) t + \varphi \right] } \sigma + e^{-i \left[ (\nu-\omega) t  +  \varphi \right]  } \sigma^\dag } . \label{HL1}
\end{eqnarray}

Second, we will write this Hamiltonian in another interaction picture with respect to the term ${\cal H}_0^{L_1} = - \hbar \Omega   \pare{ e^{ i \varphi } \sigma + e^{-i \varphi } \sigma^\dag}$. The resulting expression reads
\begin{eqnarray}
{\cal H}^{I}  \!  &=&  \! -  \frac{\hbar g }{2} \left( \left\{  \proj{+}{+} - \proj{-}{-} + e^{-i 2  \Omega t} \proj{+}{-} \right. \right. \nonumber \\
 & & -  \left.\left. e^{i 2  \Omega  t}\proj{-}{+}  \right\} e^{i \varphi} \opa + {\rm H.c.} \right) \nonumber \\  
  &-&  \frac{\hbar \lambda}{2} \left(   \left\{  \proj{+}{+} - \proj{-}{-} - e^{-i 2  \Omega  t} \proj{+}{-}  \right. \right. \nonumber \\ & & \left. \left. + e^{i 2 \Omega  t}\proj{-}{+}  \right\} e^{i (\nu-\omega) t} + {\rm H.c.} \right)   + \hbar \xi \pare{ \opa + \opa^\dag}   , \label{HI1}
\end{eqnarray}
where we have introduced the rotated spin basis $\ket{\pm} = \pare{\ket{g} \pm e^{-i \varphi} \ket{e} }/\sqrt2$. To simplify this Hamiltonian expression further, we can now set $\omega-\nu=2 \Omega ,$ and assume the amplitude of the first driving $\Omega$ to be large as compared to the rest of frequencies in~(\ref{HI1}). This implies we can apply the rotating-wave approximation (RWA), yielding the effective Hamiltonian
\begin{eqnarray}
{\cal H}_{\rm eff}
=  \frac{\hbar\lambda}{2} \sigma_z   +  \frac{\hbar g}{\sqrt2} \sigma_y \hat{p} + \hbar \xi \sqrt2 \, \hat{x} ,  \label{HamilEff}
\end{eqnarray}
where $\varphi=\pi/2$ and we make use of standard quadratures of the electromagnetic field, i.e. dimensionless $\hat{x} =(\opa+\opa^\dag)/\sqrt2$, $\hat{p} =-i(\opa-\opa^\dag)/\sqrt2$, at variance with $\hat{X}$ and $\hat{P}$ in Eq.~(\ref{Klein}), with the commutation relation $\left[ \hat{x} , \hat{p} \right] = i$. Note that $\Omega$ does not appear in the effective Hamiltonian. This is because the Hamiltonian was obtained in an interaction picture with $\Omega$ playing the role of a dominant frequency in the strong driving regime.

The Schr\"odinger equation of our system resembles that of the 1+1 Dirac equation as in Eqs.~(\ref{Dirac}) and (\ref{Klein}), where the terms $\hbar g/\sqrt2$ and $\hbar \lambda / 2$ are related to the speed of light and the mass, respectively. In addition, we have an external linear potential $ \Phi = \hbar \xi \sqrt2 \, \hat{x}$, that depends linearly on the position of the particle. Clearly, in the simulated dynamics, one can cover a wide range of physical parameters. While, for a fixed coupling strength $g$, the simulated mass is proportional to the amplitude of the weak orthogonal driving $\lambda$, the strength of the linear potential can be tuned through the amplitude $\xi$ of the longitudinal driving. This is an interesting advantage over the implementation in ion traps, where a second ion is needed to simulate the external potential \cite{KleinP,KleinPExp}. Note that in the case of a massless particle, $\lambda = 0$ and $\nu = 0$, so that $\omega = 2 \Omega$ in Eq.~(\ref{HI1}). 

As already mentioned, the study of relativistic quantum effects, such as {\it Zitterbewegung} or Klein tunneling, should be done in the phase-space representation of the electromagnetic field in the superconducting resonator. In all numerical cases studied below, the initial state of the bosonic degree of freedom of the Dirac particle is assumed to be a wavepacket with average position $\langle  \hat{x}_0 \rangle$ and average momentum $\langle  \hat{p}_0  \rangle$,
\begin{eqnarray}
\psi (x) =  \pi^{-1/4} \exp\left\{ i \langle  \hat{p}_0  \rangle  x  \right\} \,\,    \exp \left\{ -   \frac{ ( x  -  \langle  \hat{x}_0  \rangle )^2 }{2} \right\}  .    \label{psi}
\end{eqnarray}

This coincides with the $x-$quadrature representation of a coherent state of the electromagnetic field $\ket{ \frac{ \langle  \hat{x}_0  \rangle + i \langle  \hat{p}_0  \rangle }{\sqrt2} } = {\cal D} \pare{ \frac{ \langle  \hat{x}_0  \rangle + i \langle  \hat{p}_0  \rangle }{\sqrt2}  } \ket{0} $, $\ket{0}$ being the vacuum state of the bosonic field, and ${\cal D} ( \alpha ) = \exp \left\{  \alpha \opa^\dag  -  \alpha^* \opa   \right\} $ the coherent displacement operator.

For the numerical simulations discussed below, we have chosen a value of the superconducting qubit and resonator frequencies of $\omega_q=\omega=2\pi\times 9$ GHz, a qubit-field coupling of $g=2\pi\times 10$ MHz and a strong transverse driving amplitude of $\Omega=2\pi\times 200$ MHz. In addition to being experimentally realistic, this value takes into account the limitation imposed by the qubit-resonator coupling constant $g$ on the strong driving fields. The interaction time needed is of about $60$ nsec, which is well below standard decoherence times in cQED experiments.

\section{Free Dirac particle}

\begin{figure}[t]
\includegraphics[width=5cm]{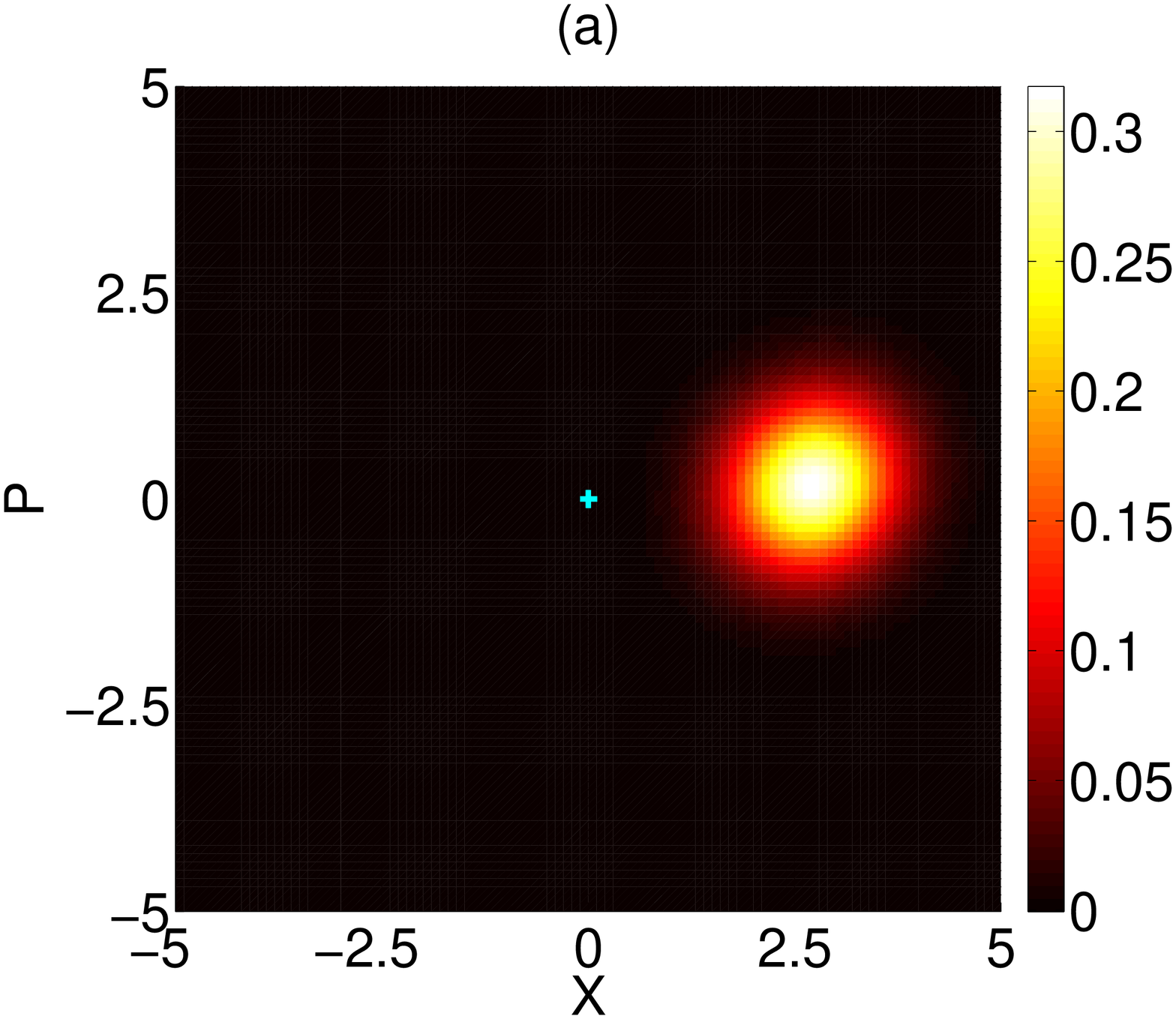}
\includegraphics[width=5cm]{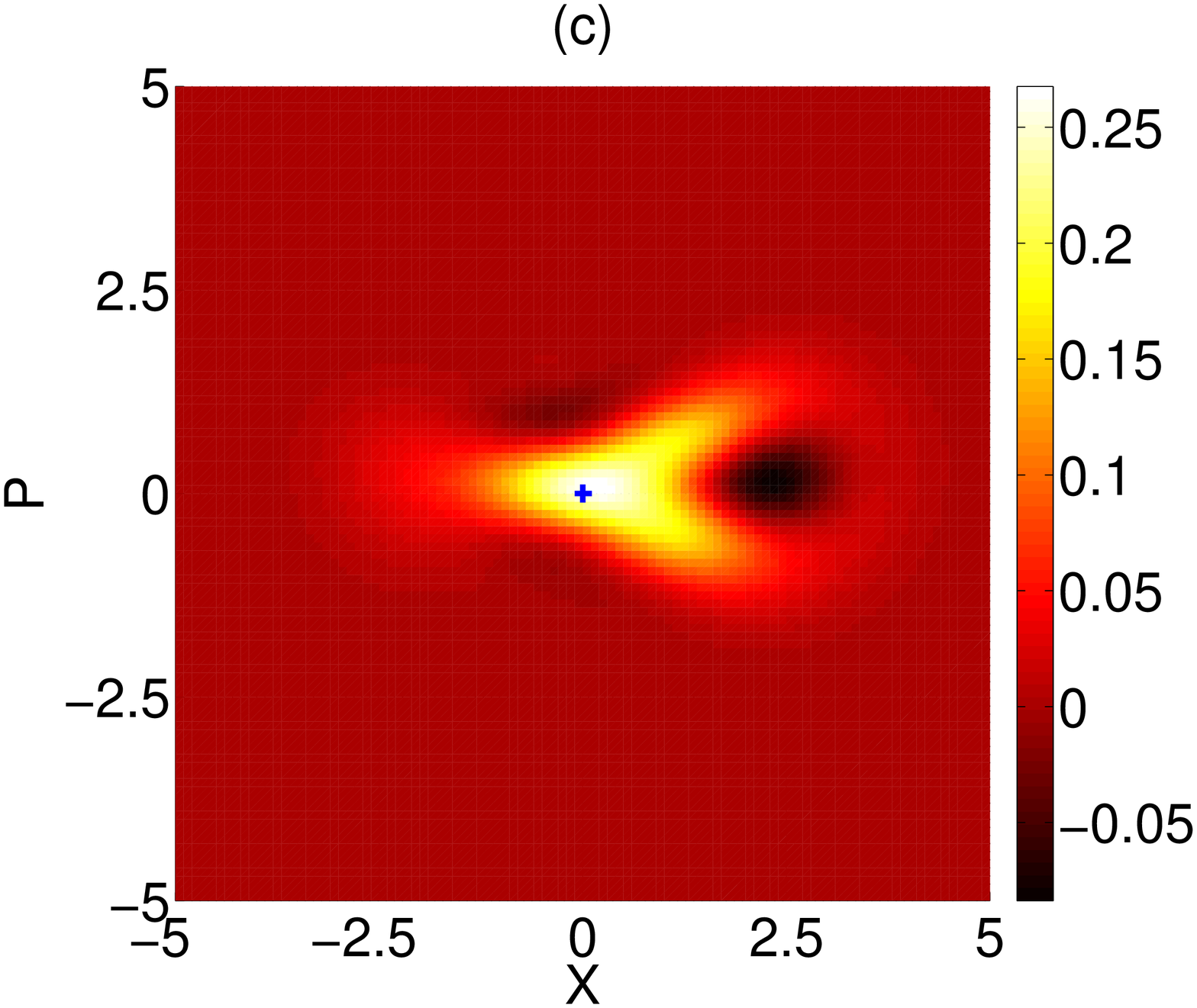}
\includegraphics[width=5cm]{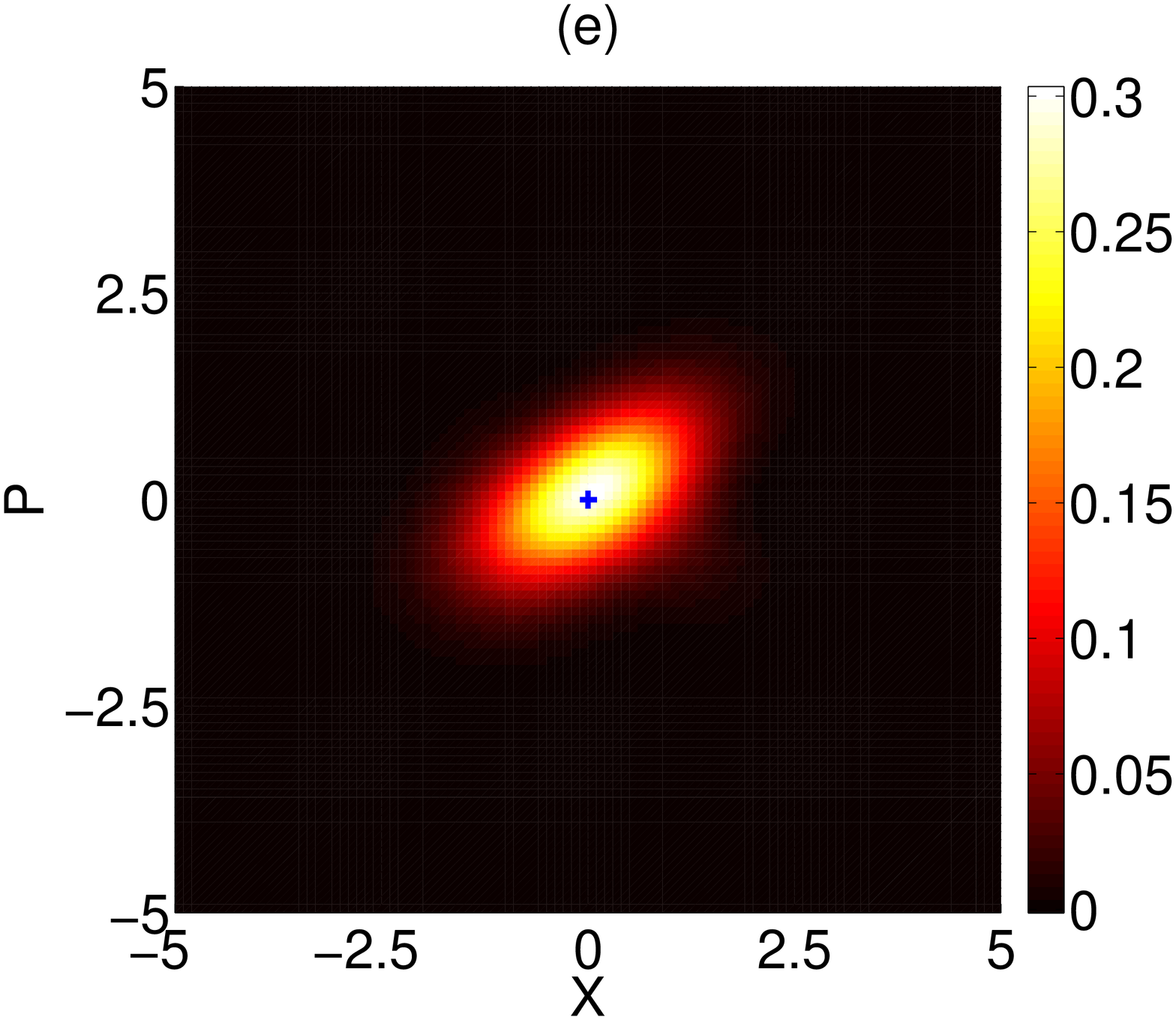} \\
\includegraphics[width=5cm]{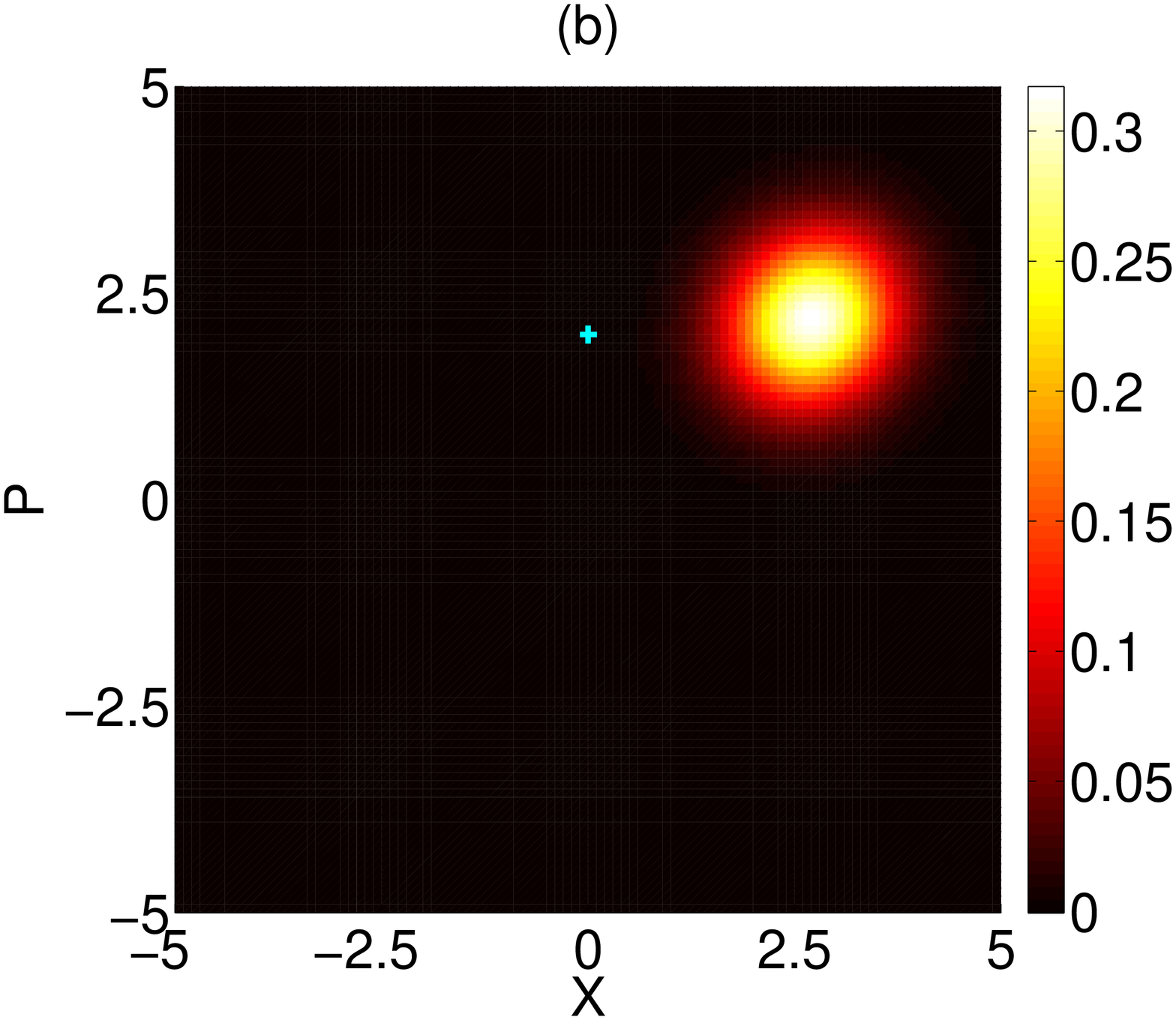}
\includegraphics[width=5cm]{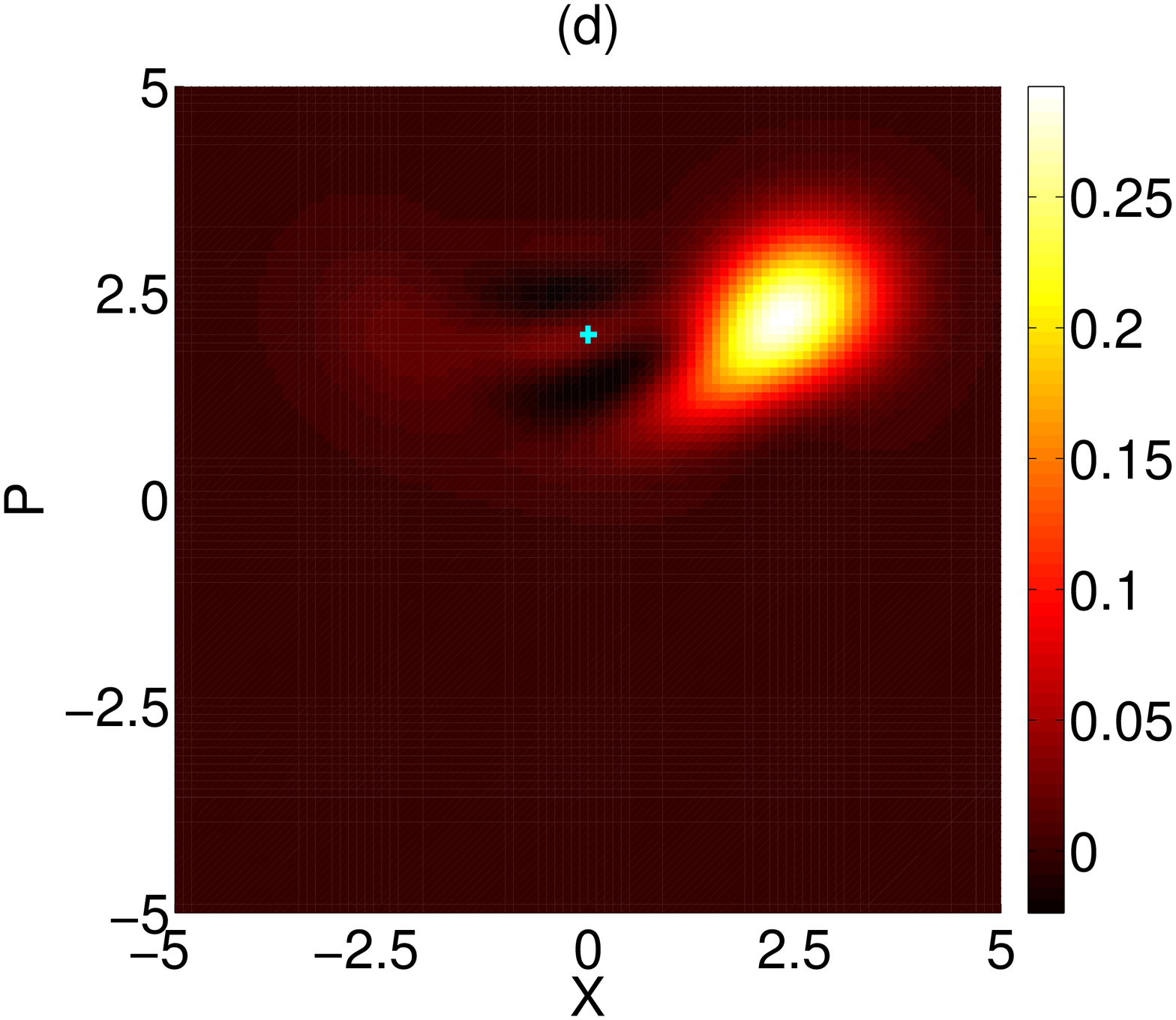}
\includegraphics[width=5cm]{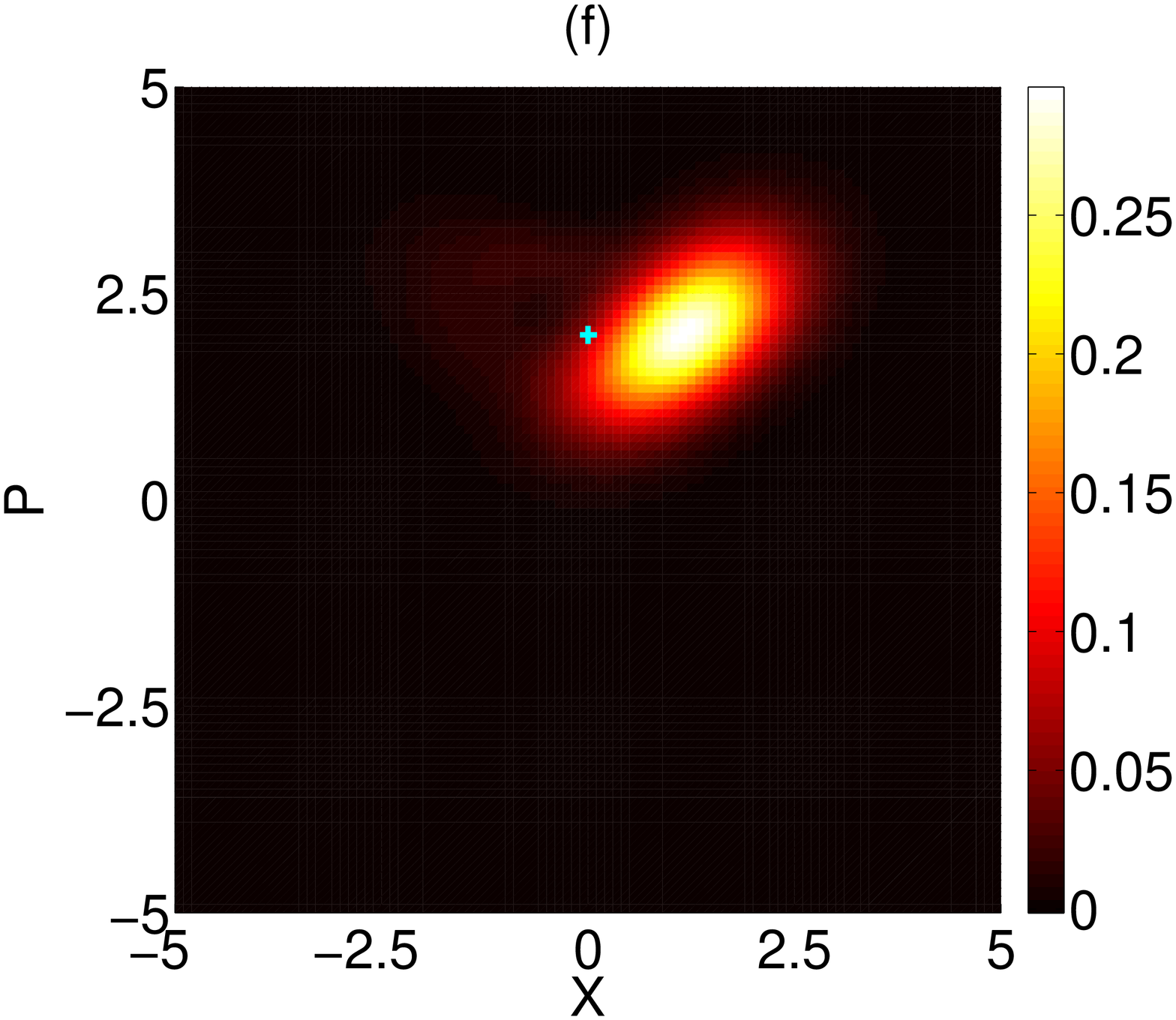}
\caption{\label{dirac-figs}Wigner function $W(x,p)$ of the field mode state inside the resonator. The initial state evolves under the Hamiltonian~(\ref{HL1}) for a time period of $60$ nsec. The physical parameters used are $g=2\pi\times 10$ MHz, $\Omega=2\pi\times 200$ MHz, $\xi=0$, together with: {\bf (a)}~$\lambda=0$ being the initial state $\ket{+,0}$; {\bf (b)} $\lambda=0$ being the initial state $\ket{+,\sqrt2i}$; {\bf (c)}~$\lambda=\sqrt2 g$ being the initial state $\ket{+,0}$; {\bf (d)} $\lambda=\sqrt2 g$ being the initial state $\ket{+,\sqrt2i}$; {\bf (e$ $)} $\lambda= 4 \sqrt2 g$ being the initial state $\ket{e,0}$; {\bf (f)} $\lambda= 4 \sqrt2 g$ being the initial state $\ket{e, \sqrt2i}$.}
\end{figure}

In the absence of external potential, $\Phi=0$, the Dirac equation for a free particle (\ref{Dirac}) does not couple the different spinor components. The Dirac Hamiltonian
\begin{eqnarray}
{\cal H}_{\rm D}=  \frac{\hbar\lambda}{2} \sigma_z   +  \frac{\hbar g}{\sqrt2} \sigma_y \hat{p}    \label{HD}
\end{eqnarray}
has two important limits depending on the value of the mass of the particle. First, in the massless case the Hamiltonian reduces to ${\cal H}_{\rm D} = \hbar g/\sqrt2  \sigma_y \hat{p}$ allowing for straightforward interpretation of its dynamics in terms of phase-space representation. Fig. \ref{dirac-figs}(a) shows the evolution of a massless particle with initial state $\ket{+,0}$. Being its spin in an eigenstate of $\sigma_y$, the effect that the interaction has is just to generate a coherent field displacement from its original position, the vacuum state $\ket{0}$, to the final one in which the state of the field is $\ket{g t /2}$. This simulates the evolution of a massless free Dirac particle moving at a speed of light stemming from $\hbar g/\sqrt2$. As we expect, the dynamics of this massless particle would be essentially unchanged if one assumes the initial state to have a non-zero value of the expectation value of momentum. This is depicted in Fig. \ref{dirac-figs}(b), where the particle starts its trajectory with $\langle  \hat{p}_0   \rangle = 2$. The particle moves exactly the same amount of space in the $x-$quadrature, ending in the coherent state $\ket{g t /2 + \sqrt2 i}$.

Second, when the mass of the particle becomes large enough, the dynamics should tend towards that of a non-relativistic one evolving under a free Schr\"odinger equation. For $\hbar \lambda/2 \gg \hbar g/\sqrt2  \langle  \hat{p}  \rangle  $, i.e. $mc^2 \gg c \langle  \hat{P}  \rangle$, one can derive a second-order effective Hamiltonian \cite{Dirac} for Eq. (\ref{HamilEff}) to yield ${\cal H}_{\rm NRel} = \sigma_z \hbar (g^2 / \lambda) \hat{p}^2$. This is indeed the nonrelativistic Schr\"odinger Hamiltonian we expected, but with the addition of the spin operator $\sigma_z$, which is reminiscent of the spinor-momentum coupling of the original Dirac equation. This spin operator ensures the existence of two decoupled equations, valid for positive and negative kinetic energies. Note that, for suitable initial states, entanglement between the spinorial and kinematic degrees of freedom is always possible, irrespective of the system being in the nonrelativistic or relativistic regimes. In Fig. \ref{dirac-figs}(e$ $), we depict the evolution of the initial state $\ket{e,0}$ under the time dependent Hamiltonian (\ref{HL1}) for the case in which the mass of the particle is such that $\hbar\lambda/2= 4\times\hbar g/\sqrt2$. This generates a squeezed vacuum state with degree of squeezing increasing linearly in time. This behaviour is in agreement with the aforementioned nonrelativistic approximation for the Hamiltonian, proportional to $\hat{p}^2$. Such a simulation realizes a standard example in all quantum mechanics textbooks. Namely, the free evolution under the nonrelativistic Schr\"odinger equation of an initial wavepacket with $\langle  \hat{p}_0   \rangle=0$ produces a particle that remains centered at the same initial position, with a wavepacket spreading over time in the $x-$quadrature.  However, the effect of the operator $\hat{p}^2$ is not simply squeezing. This can be clearly seen in Fig. \ref{dirac-figs}(f) where we have taken a particle with initial state $\ket{e,\sqrt2 i }$, that is a wavepacket with non-zero kinetic energy such that $\langle  \hat{p}_0 \rangle =2$. Now the wavepacket not only spreads over time, but its centre moves linearly in time to the right as the expectation value of the position of a nonrelativistic particle would do.

While for the limits of zero and large mass, the field state remains Gaussian during the evolution, this is no longer true when the particle mass is such that $mc^2$ becomes comparable to $c\langle  \hat{P}  \rangle$. This is the case of Figs. \ref{dirac-figs}(c) and (d), where $\hbar\lambda/2=\hbar g/\sqrt2$. Starting from initial states $\ket{+,0}$ and $\ket{+,\sqrt2i}$, respectively, the Gaussian states of the field evolve into a nonclassical ones with negative Wigner function through a highly nonlinear interaction. The effect is more evident for the case of an initial state $\ket{+,0}$ in Fig. \ref{dirac-figs}(c). Initial states with non-zero kinetic energy, such as $\ket{+,\sqrt2i }$ in Fig. \ref{dirac-figs}(d) makes the qubit-field coupling term in Eq. (\ref{HamilEff}) dominant. This interesting finding shows that a Dirac evolution generates nonclassical features mainly in intermediate relativistic regimes, assuming that one traces out over the qubit degree of freedom.

\begin{figure}
\includegraphics[width=7cm]{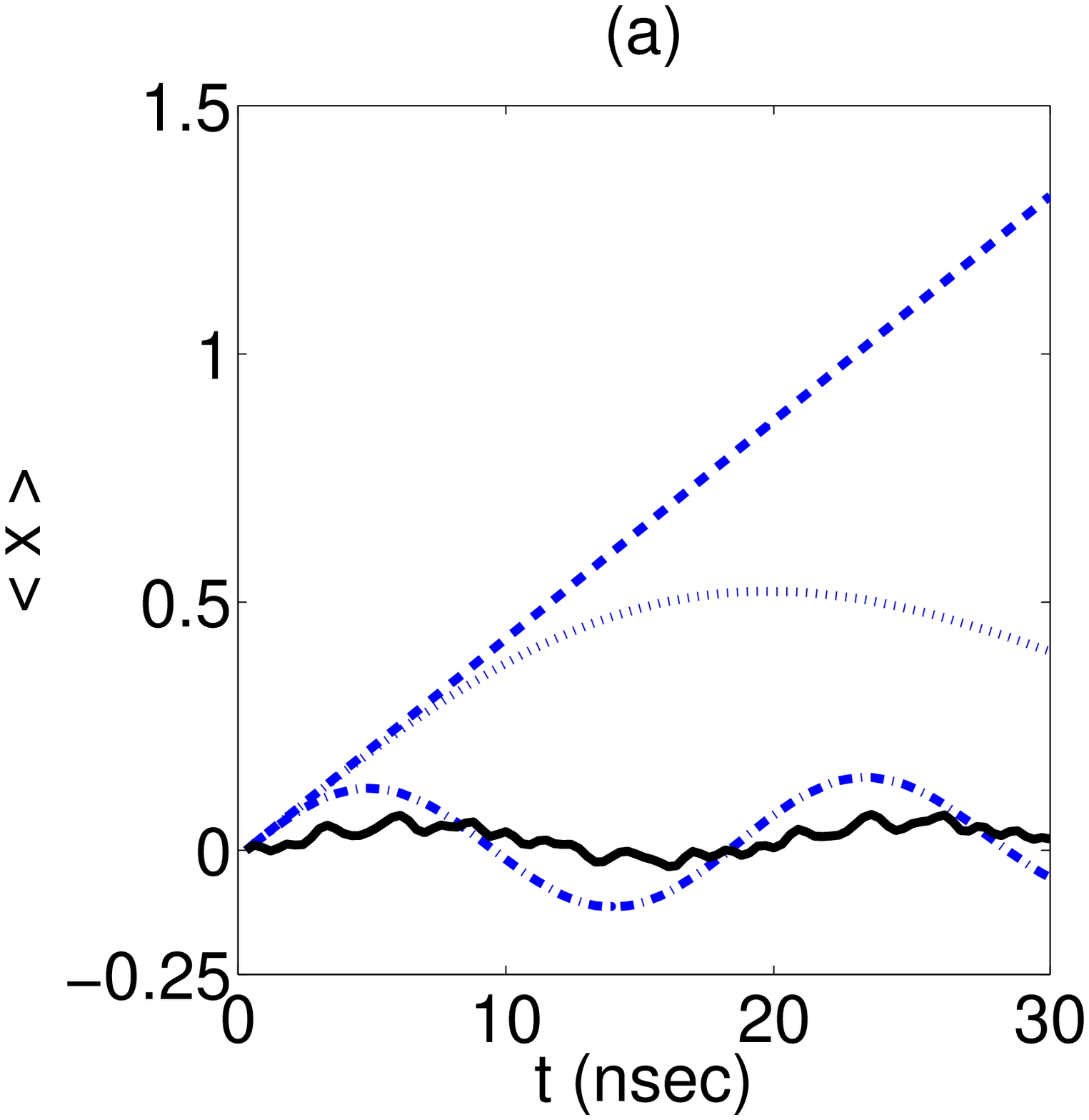} \hspace{1cm}
\includegraphics[width=7cm]{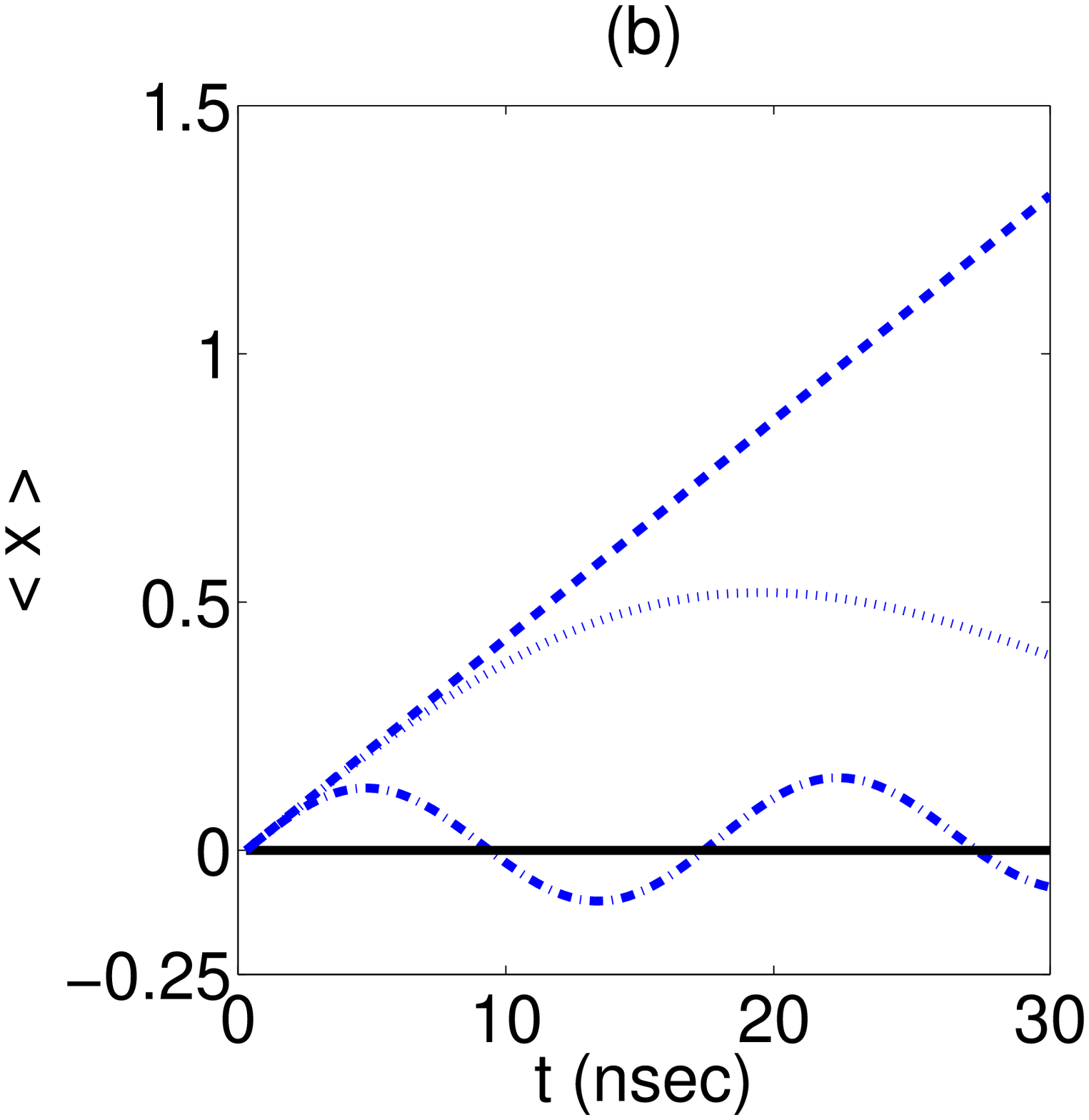}
\caption{\label{zb-figs}Expectation value of the $x-$quadrature of the electromagnetic field mode inside the resonator. The interaction time is set to $30$ nsec. The physical parameters used are $g=2\pi\times 10$ MHz, $\Omega=2\pi\times 200$ MHz, $\xi=0$, together with: $\lambda=0$ being the initial state $\ket{+,0}$ (dashed line); $\lambda=\sqrt2 g$ being the initial state $\ket{+,0}$ (dotted line); $\lambda= 4 \sqrt2 g$ being the initial state $\ket{e,0}$ (dash-dot); $\lambda= 4 \sqrt2 g$ being the initial state $\ket{e,0}$ using the FW Hamiltonian (solid line). Whereas in {\bf (a)} the calculations involve time dependent Hamiltonians (\ref{HL1}), in {\bf (b)} effective ones (\ref{HamilEff}) are used.}
\end{figure}

With our scheme we can also study the appearance of {\it Zitterbewegung} for a free spin$-1/2$ Dirac particle. In Fig. \ref{zb-figs}, we plot the time evolution of $\langle  \hat{x}  \rangle$ for a particle prepared in an initial state $\ket{+,0}$ and with three different values of the mass. For better comparison, plots (a) and (b) correspond to calculations done using the time dependent Hamiltonian (\ref{HL1}) and the effective one (\ref{HamilEff}). The good agreement found indicates that choosing a value of the strong driving amplitude $\Omega=2\pi\times 200$ MHz is enough for our purpose. The dashed line in Fig. \ref{zb-figs} shows the evolution of a massless particle, $\lambda=0$, whose expectation value of the position increases monotonically at a speed of light related to $\hbar g/\sqrt2 $. If we assume a mass such that $\hbar \lambda/2= \hbar g/ \sqrt2$, then we observe that the evolution of the $\langle  \hat{x}  \rangle$ followed by dotted line departs considerably from the constant rectilinear behaviour one would expect for a free particle. Likewise, for a mass even larger, choosing for instance $\hbar \lambda/2=4 \times \hbar g /  \sqrt2 $, the effect seen in the dash-dotted line is even more significant. Now the particle hardly moves forward. Instead, it develops an almost stationary oscillation around its initial position.

To gain in physical insight, we can derive an analytical expansion for the evolution operator of a free Dirac particle,
\begin{eqnarray}
\hspace*{-10mm} \exp\pare{-i {\cal H}_{\rm D}t/\hbar} &=& {\rm exp}\pare{-i t \sigma_y \sqrt{ \frac{\lambda^2 }{4}   +  \frac{ g^2 }{2} \hat{p}^2 } } - i \sigma_z   \frac{\lambda t}{  2}  {\rm sinc}\pare{  \sqrt{ \frac{\lambda^2 t^2}{4}   +  \frac{g^2  t^2}{2} \hat{p}^2 }  } \nonumber \\ &-& i \sigma_y \pare{   \frac{  g t \hat{p} /\sqrt2     }{ \sqrt{ \frac{\lambda^2 t^2}{4}   +  \frac{ g^2 t^2}{2} \hat{p}^2 }} - 1 }  {\rm sin}\pare{ \sqrt{ \frac{\lambda^2 t^2}{4}   +  \frac{g^2 t^2}{2} \hat{p}^2 }  }  \nonumber \\
&=& \exp\pare{-i \frac{gt}{\sqrt2}\sigma_y \hat{p}} - i \frac{\lambda t}{2} \sigma_z {\rm sinc}(gt\hat{p}/\sqrt2)  \nonumber \\
&+& \frac{\lambda^2 t^2}{4} \pare{ - \frac{ {\rm sinc}(gt\hat{p}/\sqrt2)}{2} - i \sigma_y \frac{ {\rm cos}(gt\hat{p}/\sqrt2) - {\rm sinc}(gt\hat{p}/\sqrt2)}{ gt\hat{p} \sqrt2 }  } \nonumber \\
&-& i \frac{\lambda^3 t^3}{8}   \sigma_z \frac{ {\rm cos}(gt\hat{p}/\sqrt2) - {\rm sinc}(gt\hat{p}/\sqrt2)}{ g^2t^2\hat{p}^2 } + \dots ,   \label{UDexp}
\end{eqnarray} 
where ${\rm sinc}(x)={\rm sin}(x)/x$. Here, a larger initial average momentum $g t\langle  \hat{p}(0)  \rangle /2$ leads to a faster collapse of the trembling motion for a particle prepared in a wavepacket \cite{Lurie, Zawadzki}.

As already discussed, {\it Zitterbewegung} stems from the fact that the time derivative of the position operator is not a constant of the motion for the Dirac equation. This can be seen if one writes down the evolution of this operator (\ref{evolx}). The {\it Zitterbewegung} term $\hat{Z}(t) $ will be exactly zero if the initial state of the particle contains purely positive or negative components of the spinor. However, for a completely arbitrary initial wavepacket $\hat{Z}(t) $ does not necessarily vanish. In addition, the preparation of a purely positive spinor for a massive Dirac particle is not straightforward. In Ref. \cite{DiracExp}, a heuristic method for preparing a state with only negative components was used. This shows that even for a massive Dirac particle, evolutions without {\it Zitterbewegung} are possible. Another alternative to study the existence/absence of {\it Zitterbewegung} would be to implement a measurement of the Foldy-Wouthuysen position operator \cite{Lock,FW}, 
\begin{eqnarray}
\hat{X}_{\rm FW}(t) = \hat{X}_0 + c^2 \hat{P} {\cal H}_{\rm D}^{-1} t + \hat{\Delta} ,    \label{evolFW}
\end{eqnarray}
where $\hat{\Delta}$ is the time independent operator

\begin{eqnarray}
\hat{\Delta} = \frac{ \hbar c \sigma_x }{2 \hat{E}} - \frac{  \hbar c^3 \sigma_x \hat{P}^2 }{2 \hat{E} ^2 ( \hat{E} + m c^2)} ,    \label{Delta}
\end{eqnarray}
$\sigma_x = i \sigma_z \sigma_y$, $\hat{E}=\sqrt{c^2 \hat{P}^2 + m ^2  c^4}$. The Foldy-Wouthuysen position operator in (\ref{evolFW}) has the property of being also canonically conjugate to the momentum operator. The main drawback of $\hat{X}_{\rm FW}(t)$ from an experiential point of view is that is not diagonal in the coordinate space. This means that measuring such an operator might require a full quantum tomography of the evolved state of the system.

Here, we follow a different route to the problem. Foldy and Wouthuysen are also credited for having introduced a unitary transformation \cite{FW} that allows for the direct diagonalization of the spin degree of freedom in the Dirac Hamiltonian. The physical relevance of this transformation is that it allows for a simple correspondence with classical mechanics \cite{Costella}. Using the parameters of our simulation, this new Hamiltonian is given by
\begin{eqnarray}
{\cal H}_{\rm FW} = \hat{S}_{\rm FW} \,   {\cal H}_{\rm D} \, \hat{S}_{\rm FW}^\dag   = \sigma_z  \sqrt{ \frac{\hbar^2\lambda^2}{4}   +  \frac{\hbar^2 g^2}{2} \hat{p}^2 }  \label{HFW}
\end{eqnarray} 
where
\begin{eqnarray}
 \hat{S}_{\rm FW}  = \exp \left\{ - i \, \, \,  \,  {\rm atan} \pare{ \frac{g}{\lambda\sqrt2} \sigma_x \hat{p} } \right\} . \label{SFW}
\end{eqnarray} 
If one were able to follow the evolution of the system in the Foldy-Wouthuysen representation, but without transforming the measurement apparatus, then no {\it Zitterbewegung}  would be seen. That is precisely, what we have done in the solid line of Fig. \ref{zb-figs}(b). This shows the evolution of $\langle  \hat{x}  \rangle$ for a massive particle with $\hbar \lambda/2=4\times \hbar g/\sqrt2$, just like the dash-dotted one, but using the evolution operator
\begin{eqnarray}
 {\cal U}_{\rm FW}(t)=\exp\pare{-i {\cal H}_{\rm FW}t/\hbar}=  \hat{S}_{\rm FW} \,\, \exp\pare{-i {\cal H}_{\rm D}t/\hbar}  \,\, \hat{S}_{\rm FW}^\dag \label{UFW}
\end{eqnarray} 
instead of ${\cal U}_{\rm D}(t)=\exp\pare{-i {\cal H}_{\rm D}t/\hbar}$. Now the particle remains centered at $\langle  \hat{x}  \rangle=0$ during the evolution (solid line). More importantly, for large masses we can propose a physical implementation of the Foldy-Wouthuysen transformation using the same scheme we have developed. Clearly, when $ g \langle \hat{p} \rangle/(\lambda\sqrt2)$ is small enough, we can linearize the interaction in the unitary operator $\hat{S}_{\rm FW}$. This means that
\begin{eqnarray}
 {\cal U}_{\rm FW}(t) \approx  \exp\pare{-i \frac{g }{\lambda \sqrt2} \sigma_x \hat{p }} \,\, \exp\pare{-i {\cal H}_{\rm D}t/\hbar}  \,\, \exp\pare{i \frac{g}{\lambda \sqrt2} \sigma_x \hat{p }} , \label{UFW2}
\end{eqnarray} 

where the first and last step involve the realization of a Dirac equation for a massless particle with an evolution time $t=\lambda^{-1}$. Such an implementation would require some additional local rotations and Ramsey-like driving pulses \cite{qsusc}. It would be particularly straightforward if the setup allows for a variable coupling from $-g$ to $+g$, being $g$ in strong-coupling regime. This is the case for instance, of a gradiometer type of coupling seen in flux qubits \cite{flux}. The solid line in Fig. \ref{zb-figs}(a) depicts the evolution of $\langle  \hat{x}  \rangle$, computed under these assumptions, for the same particle of large mass, with $\hbar \lambda/2=4\times \hbar  g/\sqrt2$. Even for such moderate value of the mass, it is evident how oscillations seen for the Dirac evolution become significantly reduced.

\begin{figure}[t]
\includegraphics[width=5cm]{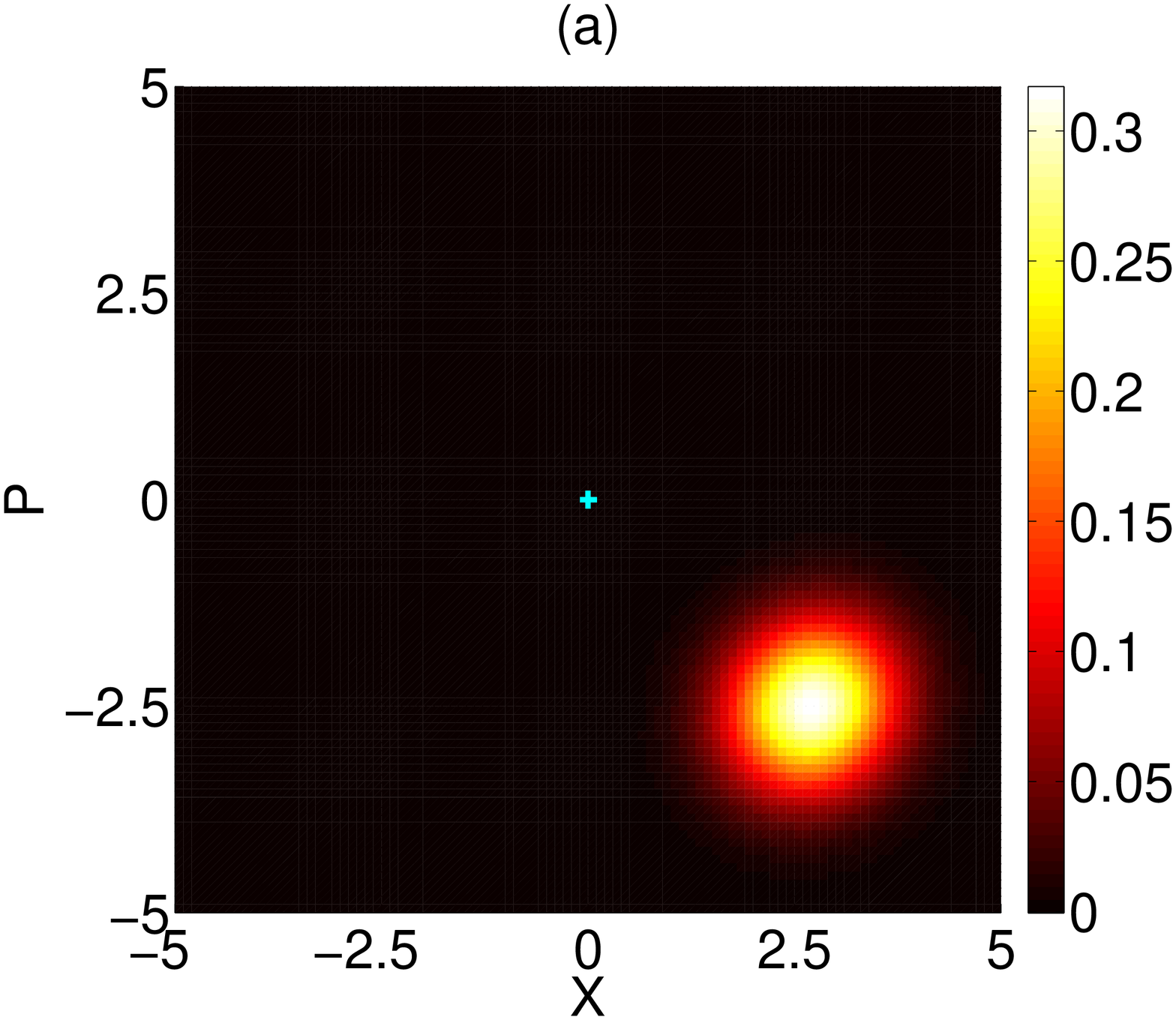}
\includegraphics[width=5cm]{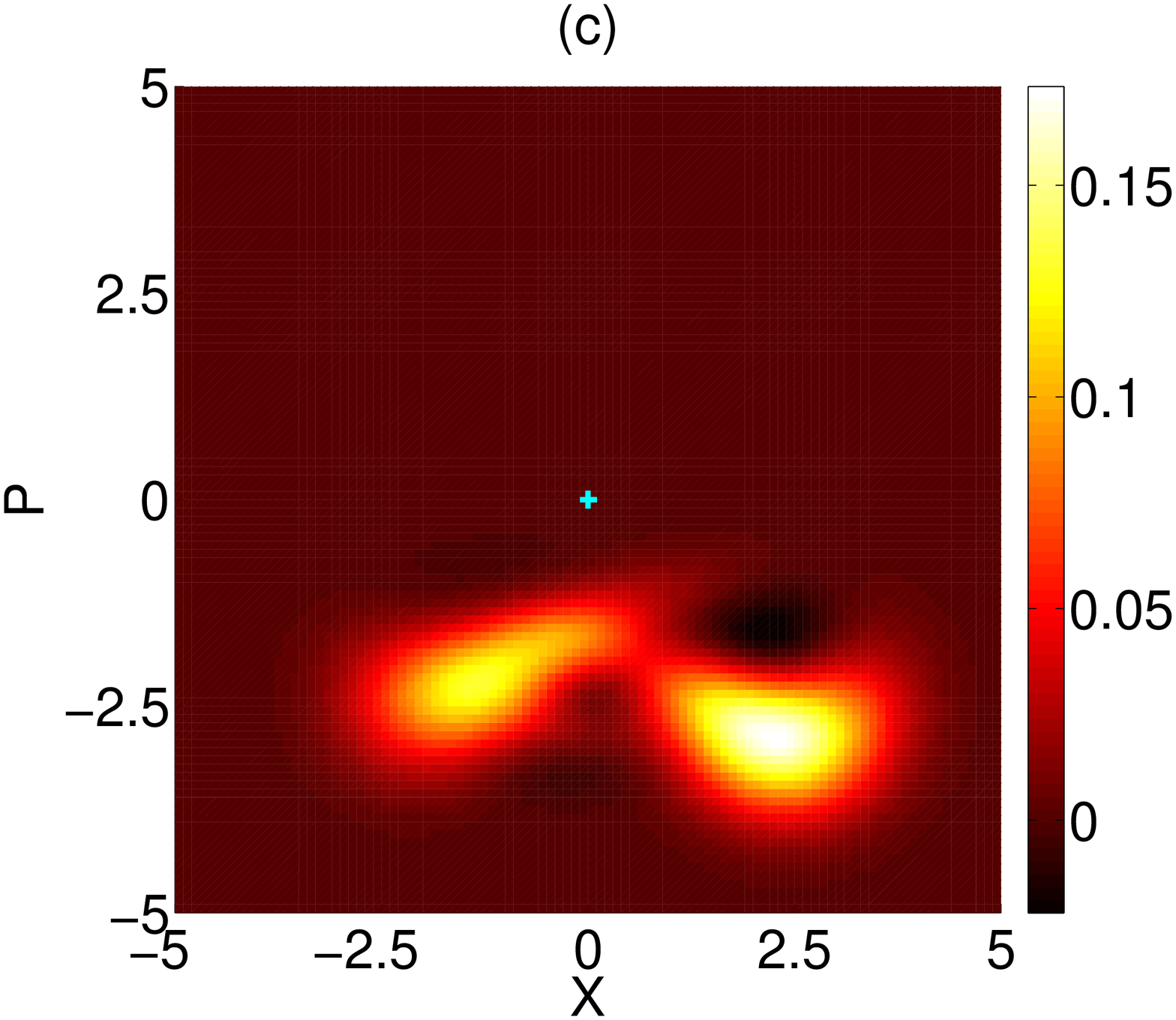}
\includegraphics[width=5cm]{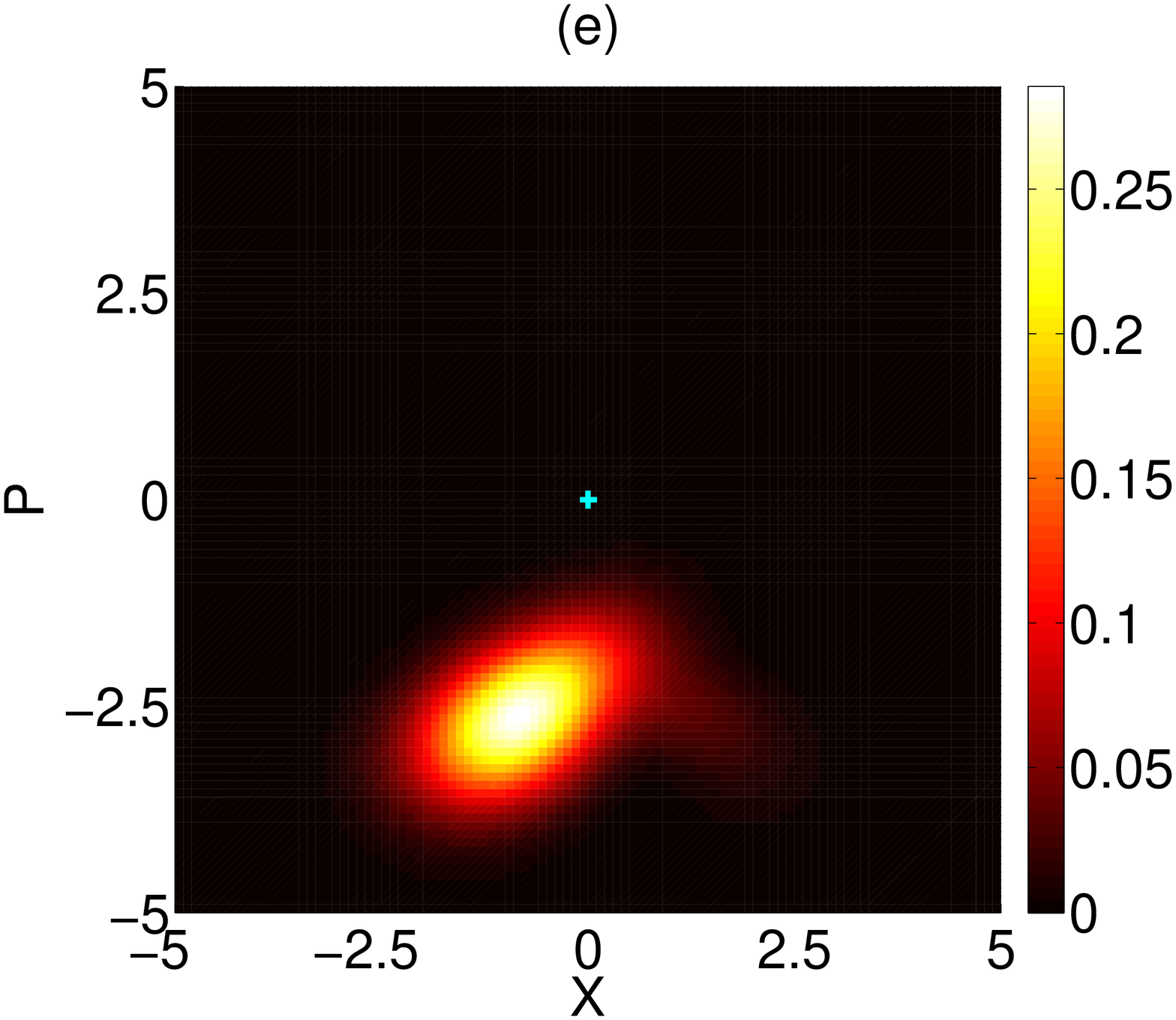} \\
\includegraphics[width=5cm]{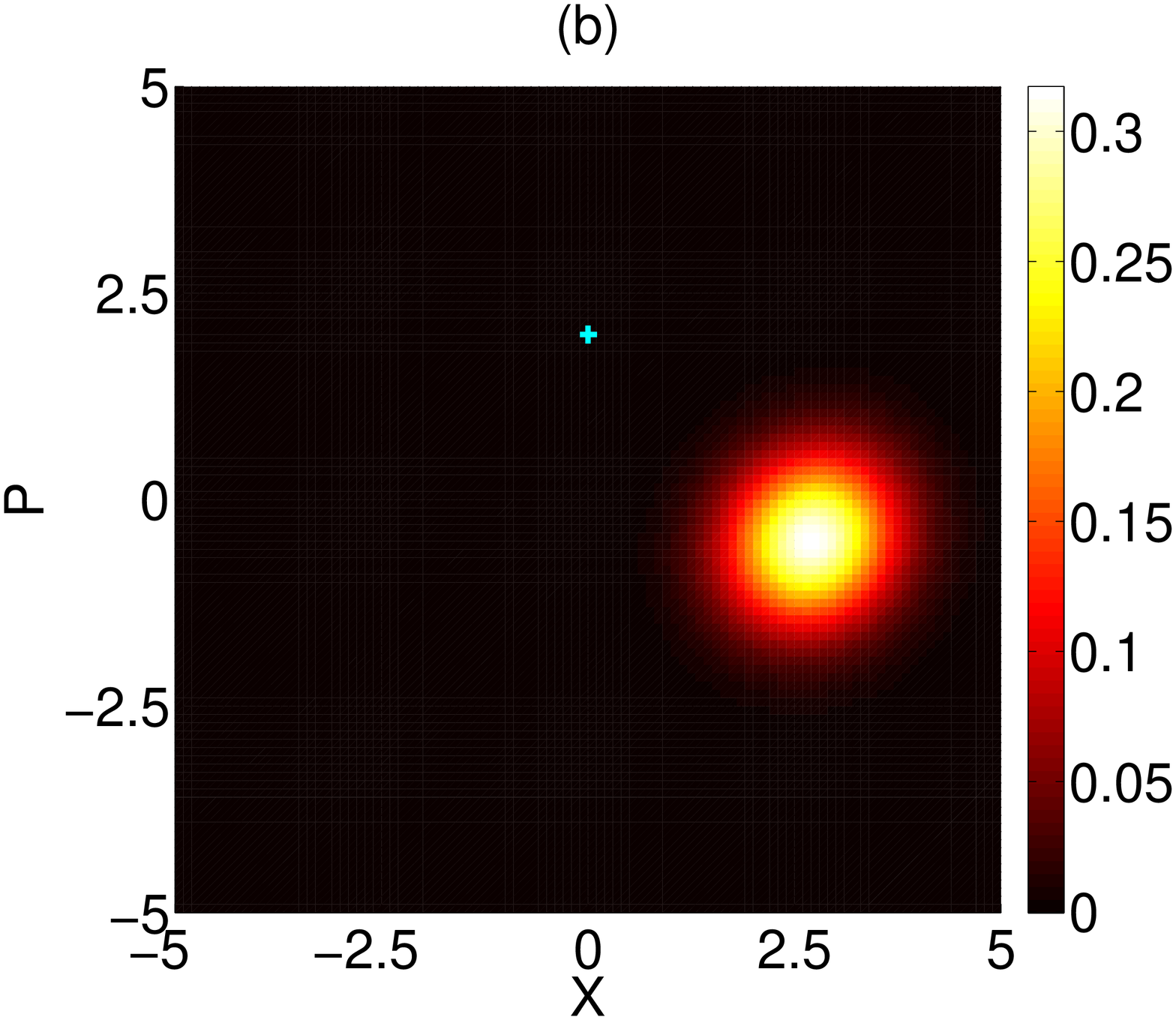}
\includegraphics[width=5cm]{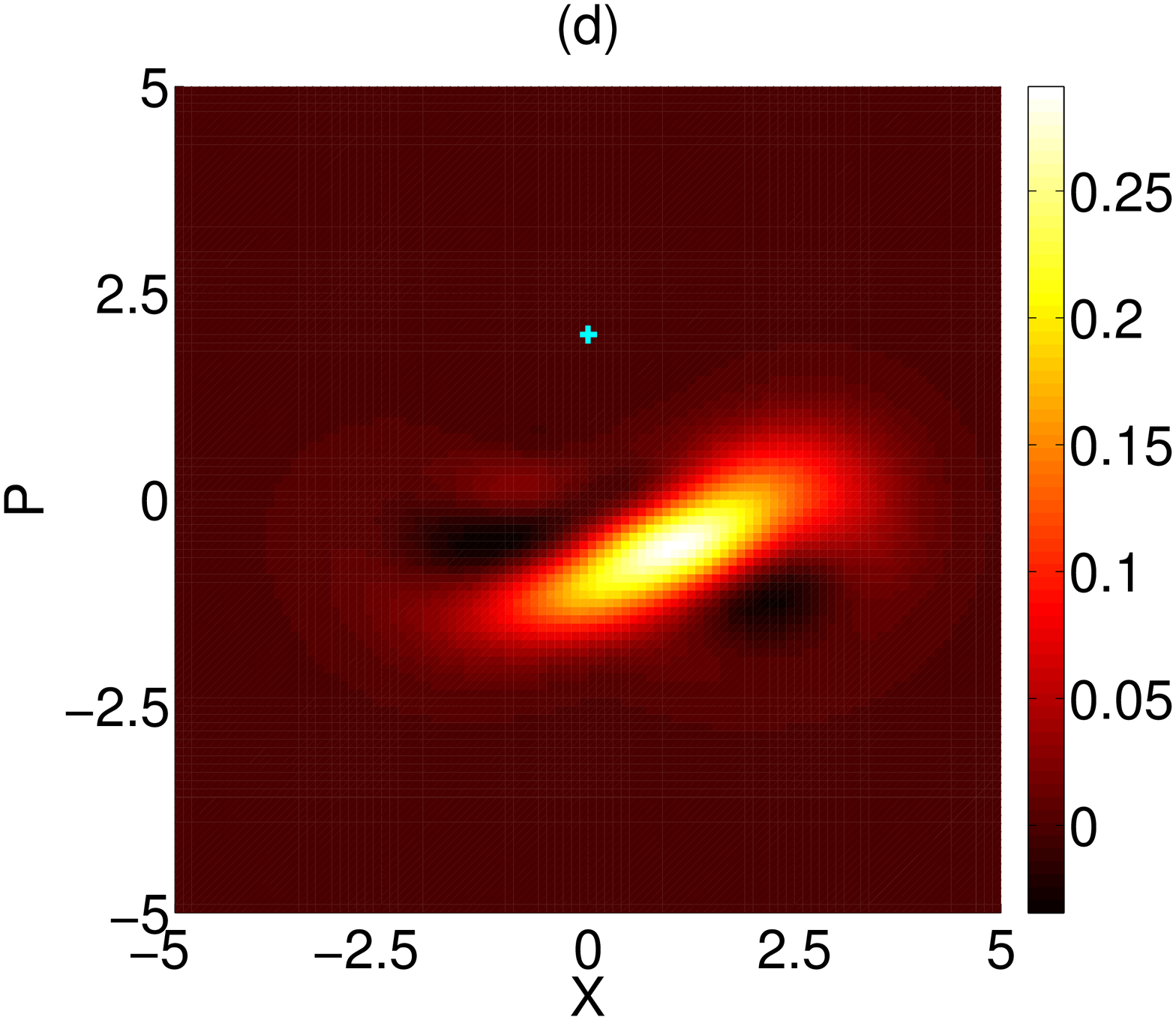}
\includegraphics[width=5cm]{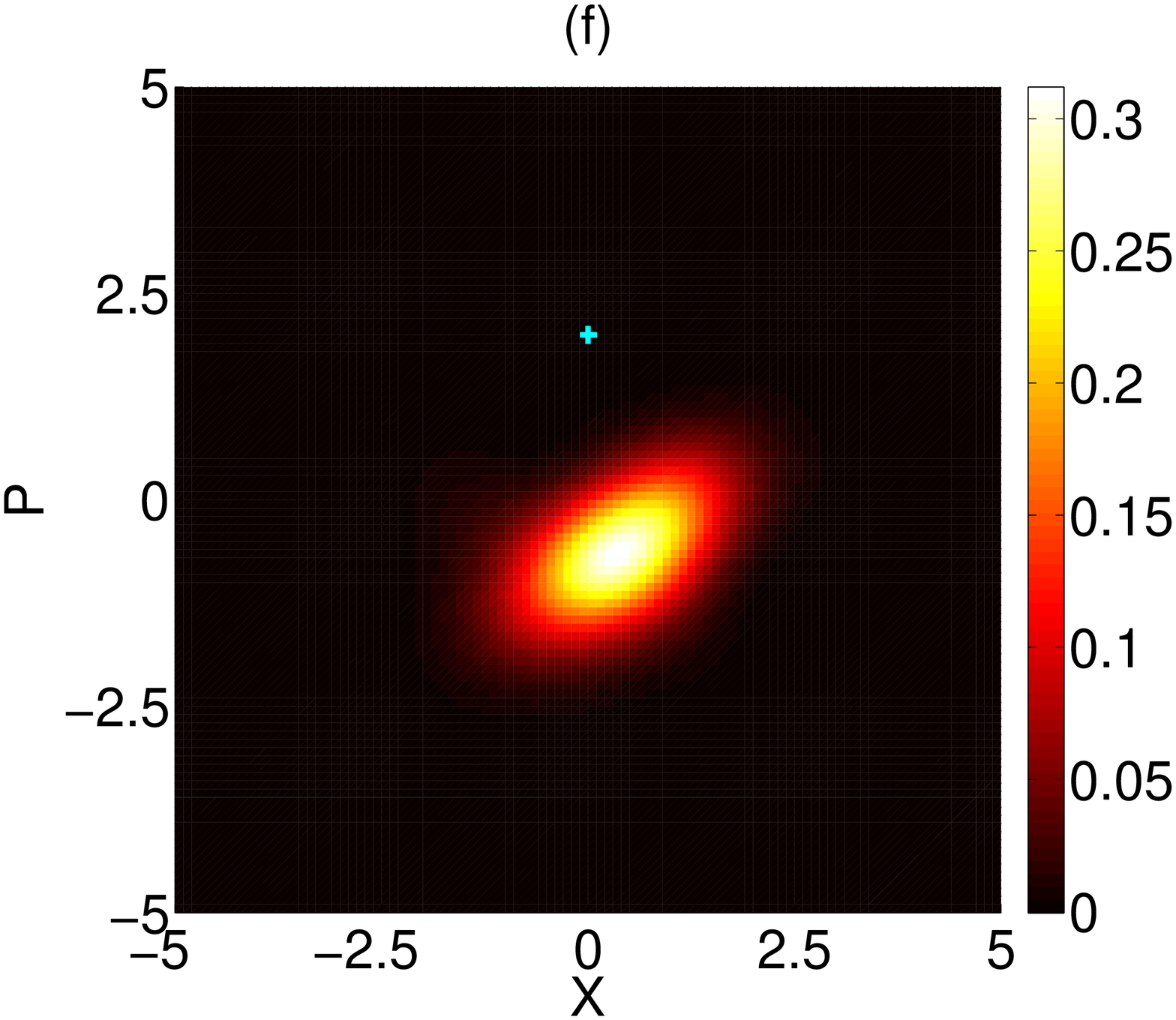}
\caption{\label{klein-figs}Wigner function $W(x,p)$ representation of the state of the electromagnetic field mode inside the resonator. The initial state evolves under the Hamiltonian (\ref{HL1}) for a time $60$ nsec. The physical parameters used are $g=2\pi\times 10$ MHz, $\Omega=2\pi\times 200$ MHz, $\xi=g/2$ and together with: {\bf (a)} $\lambda=0$ being the initial state $\ket{+,0}$; {\bf (b)} $\lambda=0$ being the initial state $\ket{+,\sqrt2i}$; {\bf (c)} $\lambda=\sqrt2 g$ being the initial state $\ket{+,0}$; {\bf (d)} $\lambda=\sqrt2 g$ being the initial state $\ket{+,\sqrt2i}$; {\bf (e)} $\lambda= 4 \sqrt2 g$ being the initial state $\ket{e,0}$; {\bf (f)} $\lambda= 4 \sqrt2 g$ being the initial state $\ket{e, \sqrt2i}$.}
\end{figure}

\section{Dirac particle in an external potential: Klein tunneling}

The addition of an external potential has the effect of coupling the different energy components of the spinor. As before, in the evolution under an external potential we can identify two limiting cases. A massless Dirac particle is described by the Hamiltonian ${\cal H}_{\rm K} =    \hbar g/\sqrt2  \sigma_y \hat{p} + \hbar\xi\sqrt2 \, \hat{x}$. In Figs. \ref{klein-figs} (a) and (b) the evolution of the initial state $\ket{+,0}$ and  $\ket{+,\sqrt2i}$, respectively, is shown. We use the same set of parameters as in Fig. \ref{dirac-figs}, with the only addition of a linear potential of strength $\xi=g/2$. Again, starting with the qubit state, $\ket{+}$, allows for a simple interpretation. The initial state of the field remains coherent while being subject to two independent displacements. Namely, as for the free-particle case, there is one in the $x-$quadrature proportional to $g/\sqrt2$, and the second one induced by the potential that drags it down in the $p-$quadrature. Hence, the existence of such an external potential cannot alter the rectilinear movement in position representation at the speed of light. This is precisely the effect of the Klein paradox. A massless Dirac particle may propagate through the potential barrier with finite probability, regardless of the potential strength. The tunneling is accompanied by the 'rotation' of the positive energy components of the initial spinor. In our simple case, this means that the components with positive momentum components in the Gaussian wavepacket will be pushed downwards by the potential becoming negative. And the rate at which this happens is given by $\xi \sqrt2$. This explains why the only noticeable difference between the plots in Figs. \ref{dirac-figs} and \ref{klein-figs} is in the value of $\langle  \hat{p}  (t) \rangle$.
 
If the mass of the particle is large enough, then the nonrelativistic Schr\"odinger equation should be a good approximation. Its Hamiltonian can be written as ${\cal H}_{\rm NRel} = \hbar \sigma_z \hat{p}^2/\lambda + \hbar\xi \sqrt2 \, \hat{x}  $, which guarantees that any initial Gaussian state should remain Gaussian during the evolution. This is indeed what we see in Figs. \ref{klein-figs}(e$ $) and (f) where we have prepared initial states $\ket{e,0}$ and $\ket{e,\sqrt2i}$, respectively, with a mass such that $\hbar \lambda/2=4\times \hbar g/\sqrt2$. In the first of them, the particle is being pushed backwards by the potential. The same can be said for the second. Although having an initial positive kinetic energy has allowed it to enter the barrier further, after $60$ nsec it is already moving backwards.

While for these two limiting cases we see complete transmission or reflection, a particle with an intermediate mass will present only partial transmission/reflection. This is shown in Fig. \ref{klein-figs}(c) which corresponds to the initial state $\ket{+,0}$ to a massive particle with $\hbar \lambda/2=  \hbar g/\sqrt2$. The wave packet has now broken up into spinor components of different sign which move away from the center of the potential. If the wavepacket has some initial kinetic energy, as in Fig. \ref{klein-figs}(d) where we prepare $\ket{+,\sqrt2i}$, then the particle tunnels slightly to stop and break up eventually.

As we did in Eq. (\ref{UDexp}), one can obtain an analytical expansion for the evolution operator of a Dirac particle under the effect of a linear potential. Using the formula
\begin{eqnarray}
\hspace*{-10mm} \exp\left\{-i \pare{ \frac{\lambda t}{2} \sigma_z   +  \frac{ g t }{2} \sigma_y \hat{p} + \xi \sqrt2 \,t\,  \hat{x} }  \right\}  \nonumber \\ =  \exp\pare{-i \frac{{\cal H}_{\rm D}}{\hbar} t }  \exp\pare{-i \xi \sqrt2 \,t\,  \hat{x} } \exp\pare{i \frac{ g \xi t^2}{2}  \sigma_y}  + O\pare{t^3 } ,   \label{UKexp}
\end{eqnarray} 
we see that for short time intervals the evolution corresponds to that of the free Dirac particle (\ref{UDexp}), but with an initial state with rotated spin and displaced initial momentum with respect to the prepared wavepacket.

The reconstruction of the Wigner function will allow to characterise what type of dynamics is being simulated. This can be done either using intracavity techniques \cite{Martinis} or allowing the quantum field to leak the superconducting resonator for its subsequent measurement. This will require the use of methods for reconstructing the moments of a propagating quantum microwave signal \cite{Menzel,Wallraff,Itinerant}, already available in cQED.

\section{Conclusions}

We have shown that circuit QED provides a powerful platform for simulating the relativistic quantum physics, establishing a useful dialogue between unconnected fields. Tuning the parameters of three classical microwave drives, the system is made to evolve according to the Dirac equation for a particle in presence of a linear potential. The degree of controllability offered by this scheme would allow to study interesting physical phenomena far beyond what any direct experiment has achieved. A relevant example is the possibility of implementing the Foldy-Wouthuysen transformation, which for decades has remained a purely abstract construction.

\section*{Acknowledgments}
We thank Leo DiCarlo, Jorge Casanova, and Lucas Lamata, for insightful discussions. The authors acknowledge funding from EPSRC EP/H050434/1, Spanish MINECO FIS2012-36673-C03-02; UPV/EHU UFI 11/55; UPV/EHU PhD fellowship; Basque Government IT472-10; SOLID, CCQED, PROMISCE, and SCALEQIT European projects.

\end{document}